\documentclass[aps,prb,superscriptaddress,showpacs,amsmath,twocolumn]{revtex4-1}

\usepackage{amsmath}
\usepackage{amssymb}
\usepackage{epsfig}
\usepackage{subfigure}

\newcommand{\igr}[2][]{\includegraphics[#1]{#2}}

\begin{document}

\author{Csaba G. P\'eterfalvi}

\author{Colin J. Lambert}
\email{c.lambert@lancaster.ac.uk}
\affiliation{Department of Physics, Lancaster University, Lancaster, LA1 4YB, UK}

\title{Suppression of single-molecule-conductance fluctuations using extended anchor groups on graphene and carbon-nanotube electrodes}

\begin{abstract}
Devices formed from single molecules attached to noble-metal electrodes exhibit large conductance fluctuations, which inhibit their development as reproducible functional units. We demonstrate that single molecules with planar anchor groups attached carbon-based electrodes are more resilient to atomic-scale variation in the contacts and exhibit significantly-lower conductance fluctuations. We examine the conductance of a 2,6-dibenzylamino core-substituted naphthalenediimide (NDI) chromophore attached to carbon electrodes by either phenanthrene anchors or by more extended anchor groups, which include OPE spacers. We demonstrate for the more spatially-extended anchor groups, conductance fluctuations are significantly reduced. The current-voltage characteristic arising from long-range tunnelling, is found to be strongly non-linear with pronounced conductance suppression below a threshold voltage of approximately 2.5 volts.
\end{abstract}

\pacs{31.15.ae, 31.15.at, 31.15.es, 71.15.Mb, 73.23.Ad, 73.63.Fg, 73.63.Rt}

\maketitle

Combined experimental and theoretical studies have provided new
insights into the interplay of molecular conformation, electronic
structure and electrical conductance.
\cite{g1} in single-molecule electronic devices. It is clear that measured conductance values depend on the atomic-scale contact geometry of the electrodes,
\cite{g2} temperature,
\cite{g3} the local environment of the system (vacuum or air, solvent, \textit{etc.})
\cite{g4} and molecular features such as the extent of conjugation,~\cite{salomon_comparison_2003,kushmerick_effect_2002} the nature of the terminal anchor groups (\textit{e.g.} thiol, amine, carboxylic acid),~\cite{li_conductance_2006} the detailed conformation
\cite{g8} and tilt angle
\cite{g9} of the molecule in the junction, and on quantum
coherence and interference of electrons transiting the molecule
\cite{sedghi_long-range_2011,g11}.

With a view to developing stable sub-10nm electronic devices, a
great deal of effort has been devoted to understanding the
interplay between terminal anchor groups and metallic electrodes.
The ideal molecular anchor group should form reproducible and
mechanically-stable contacts with well-defined binding sites and
strong electronic coupling to the nanoelectrodes.~\cite{g12}
Recently, chemical synthesis was used to vary the anchor groups
within a family of molecules and the relative anchoring
performance of four different terminal groups (SH, pyridyl (PY),
NH2 and CN) was measured and calculated. This study revealed the
following sequence for junction formation probability and
stability: \mbox{PY $>$ SH $>$ NH2 $>$
CN.~\cite{hong_single_2011}} In common with other studies of
single molecules attached to gold electrodes, broad variations in
the measured conductances are observed, which suggests that
alternative electrodes to gold are needed, if reliable and
reproducible devices are to be
developed.~\cite{salomon_comparison_2003} Alternatives based on
surface grafting \textit{via} covalent bonds, such as
carbon-carbon,~\cite{ranganathan_covalently_2001}
metal-carbon~\cite{cheng_situ_2011} and silicon-carbon~\cite{g16}
also exhibit significant sample-to-sample fluctuations.

In this paper, as a possible route to reducing fluctuations, we
examine the use of extended planar anchor groups attached
\textit{via} $\pi-\pi$ interactions to either graphene or carbon
nanotube (CNT) electrodes.  Recent experiments suggest that
carbon-based electrodes are a viable alternative to the more
commonly-used noble metals~\cite{g17,prins_room-temperature_2011}.
For example, Prins \textit{et al}~\cite{prins_room-temperature_2011} created a few-nm-wide gap in graphene, \textit{via} feedback controlled electroburning and carried out room temperature, gate controlled measurements on a single molecule bridging the gap. Similarly Marquardt \textit{et al}~\cite{marquardt_electroluminescence_2010} used electrical breakdown to form nanogaps in free-standing CNTs and bridged the gap with a single molecule.

Nanogap electrodes of this kind have several advantages over more-commonly used noble-metal electrodes, because the absence of screening makes it easier to gate the molecule and their planar structure makes it possible to image the molecule in situ. In this paper, we perform simulations based on density-functional theory (DFT) to demonstrate that such carbon-based electrodes have the further advantage of allowing the suppression of conductance fluctuations arising from atomic-scale disorder by increasing the size of planar anchor groups.

\begin{figure}[ht!]
\centering
  \igr[width=\linewidth,trim=105 0 95 0,clip]{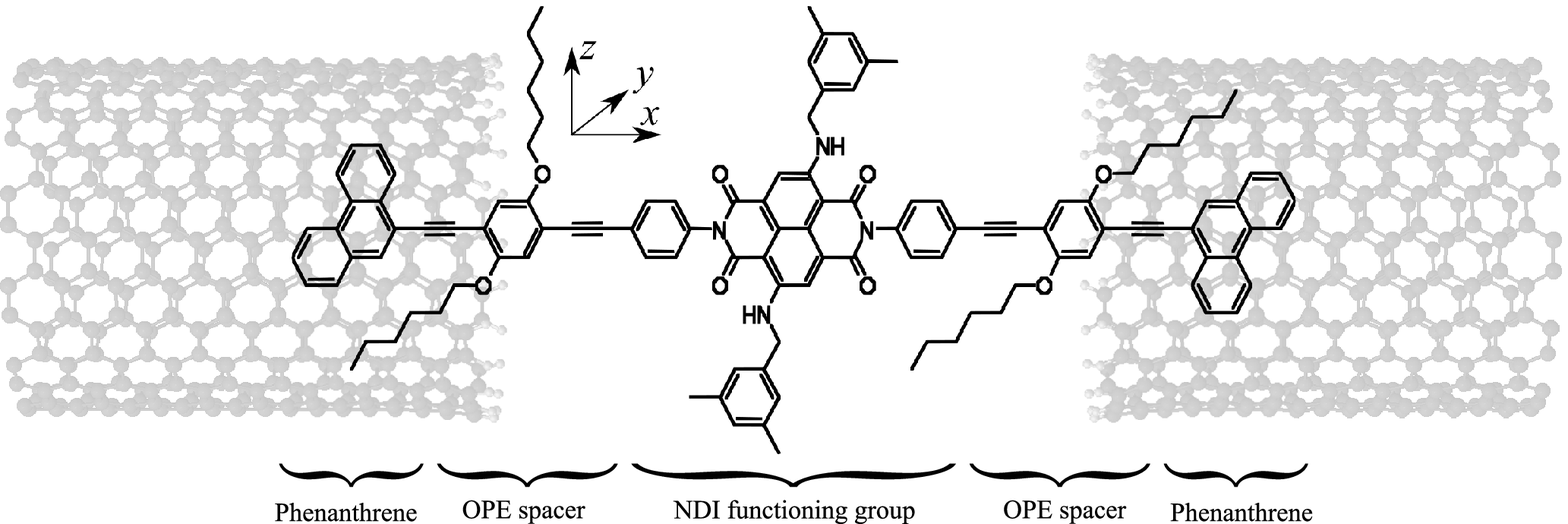}
  \caption{The studied molecule with an NDI functioning group, OPE spacers and Phenanthrene anchor groups on the top of the CNT junction (shown in light grey).}
  \label{fig:molecule}
\end{figure}

As a specific example, we investigate the electronic properties of
a rod-like molecule, closely related to that measured by Marquardt
\textit{et
al}~\cite{marquardt_electroluminescence_2010,blaszczyk_synthesis_2006},
which has a 2,6-dibenzylamino core-substituted naphthalenediimide
(NDI) functional group, connected \textit{via} oligophenylene
ethynylene (OPE) linkers to phenanthrene terminal groups, as shown
in Fig.~\ref{fig:molecule}. This contains the same NDI core as
that measured by Marquardt \textit{et al}, but theirs contains
three OPEs in the left and right linkers, whereas the molecule in
Fig.~\ref{fig:molecule} contains only a single OPE in each linker.
The longer 3-3 OPE molecule of Ref.~\onlinecite{marquardt_electroluminescence_2010}, when
attached to electrodes, is prohibitively expensive to simulate,
whereas the 1-1 OPE of Fig.~\ref{fig:molecule} is more tractable.

To model carbon-based electrodes, we couple the planar end groups
of the molecule of fig. 1 to a large-diameter $(12,12)$
single-walled, metallic armchair CNT-electrodes, \textit{via}
$\pi$ overlap, as shown in Fig.~\ref{fig:molecule}. Since the CNTs
are locally flat on the scale of the phenanthrene terminal groups
and OPE bridge, our results also capture generic transport
properties obtained using alternative carbon-based electrodes,
such as multi-wall CNTs and mono- or few-layer graphene.

Our first goal is to compute the $I-V$ characteristics of this molecule, and show that the threshold voltage measured by Marquardt \textit{et al} is an intrinsic
property of this class of molecules. Our second goal is to show that by increasing the size of the planar anchor groups, the sensitivity to lattice defects can be reduced.

\section{Theoretical method}

The structure shown in Fig.~\ref{fig:molecule} involves over 1100 atoms and therefore an efficient computational scheme is needed to determine its electrical properties. The first step is to calculate the relaxed geometry, for which we use the DFT code SIESTA.~\cite{soler_siesta_2002} A double-zeta polarized basis set is chosen, the exchange correlation is described by GGA~\cite{perdew_generalized_1996} and the atoms are relaxed until all forces are less than 0.05\,eV/{\AA}. The size of our molecules prevents the inclusion of van der Waals interaction within a currently-tractable calculation. Nevertheless it is known that van der Waals interactions tend to increase the binding energy of smaller molecules to electrodes, and the effect on transmission is to make the anchor groups more transparent. In our case however, the transmission is dominated by tunnelling through the HOMO-LUMO gap and therefore increasing the transparency of the contacts is not likely to produce a significant change. The relaxation consists of several steps: the CNT leads and the molecule are relaxed separately, and then the whole system again relaxed. After introducing lattice defects into the once-relaxed system, the structure is again relaxed. For consistency, the same force tolerance is used in all cases.

For the purpose of structure relaxation, the system is
infinitely-periodic along $x$-axis and for the small (large) end
groups, comprises nanotube sequences between the molecules made up
of 17 (21) CNT unit cells. These are long enough to ensure
convergence of the mean-field DFT Hamiltonian parameters, which
are later employed to compute transmission coefficients. On the
surfaces of the nanogaps, the $(12,12)$ armchair CNTs are
terminated with hydrogen atoms to saturate the dangling bonds.

After relaxing the whole periodic system, the underlying DFT tight-binding Hamiltonian is extracted, within a double-zeta basis-set representation and  the transmission coefficient $T(E)$ is calculated using an equilibrium implementation of the \textit{ab-initio} transport code SMEAGOL,~\cite{rocha_spin_2006} which is based on a Greens-function scattering technique. At temperature $T$, and bias $V$, the current $I$ can be calculated using the integral:
\begin{equation*}
I(V)=\frac{2e}{h}\int_{-\infty}^{\infty} \left(f(E-eV/2)-f(E+eV/2)\right) T(E) \textrm{d} E\ ,
\end{equation*}
where $f(E)$ is the Fermi function.

\section{Lattice defect resilience}

To show that extended anchor groups make the system resilient to lattice defects, we introduce defects in the lattice of the CNT electrodes mainly around the anchor
groups, and allow the bulk of the CNT to remain pristine. We introduce defects in two stages: first, we remove 6-6 carbon atoms from the edges, and introduce 1-1
{\it Stone--Wales defects}~\cite{g24}
and 1-1 vacancies in contact region of each electrode, which
consists of the last couple of unit cells in the vicinity of the
end of the CNTs.
\begin{figure}[!ht]
\centering
    \subfigure[\small\ Small contacts with few defects]{\label(SubFig1){subfig:jsf}
    \igr[height=63pt]{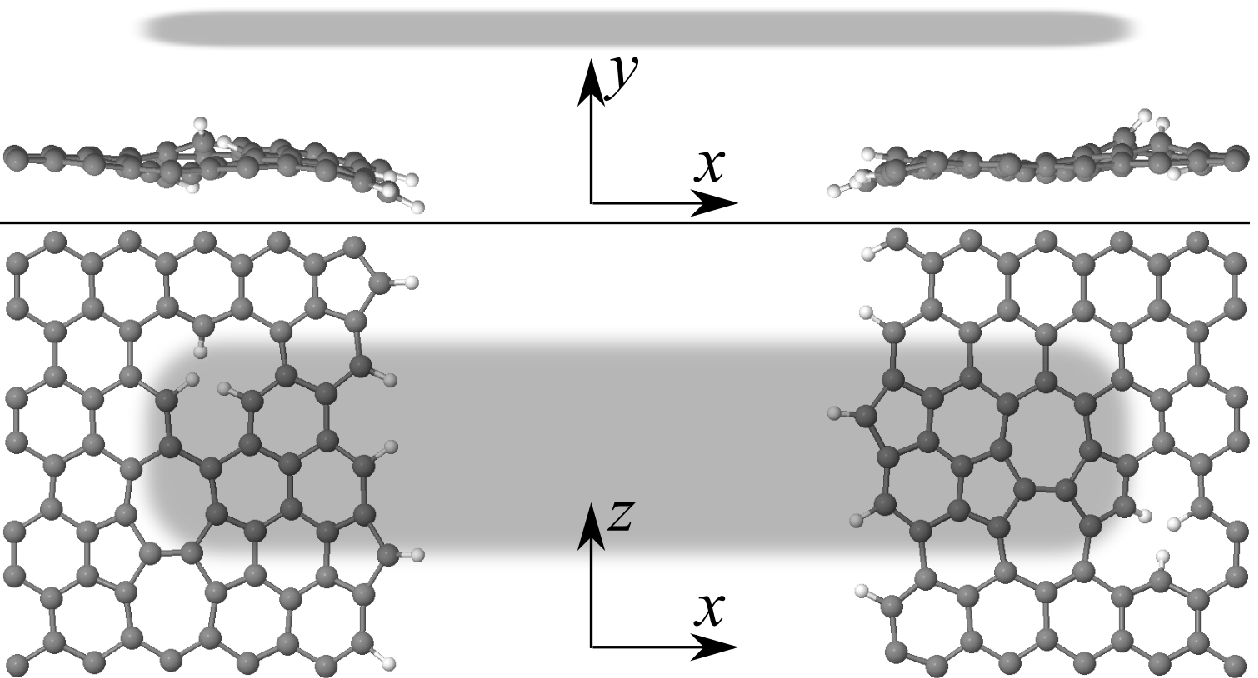}}
    \subfigure[\small\ Large contacts with few defects]{\label(SubFig1){subfig:jlf}
    \igr[height=63pt]{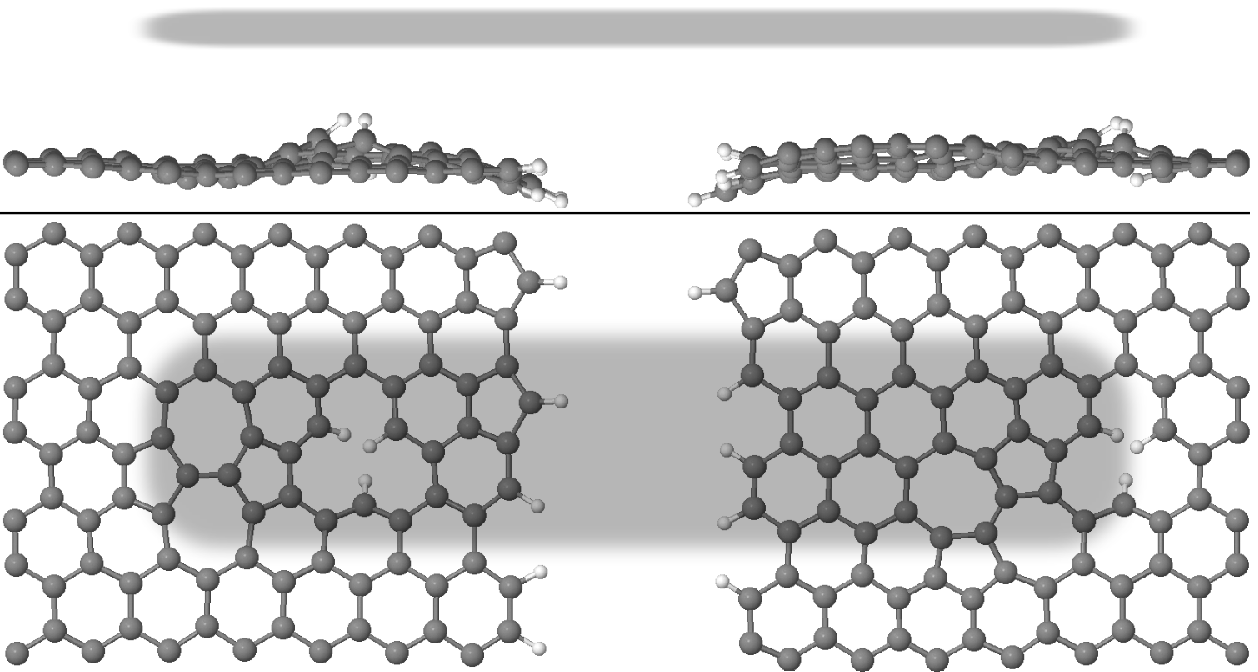}}
\\
    \subfigure[\small\ Small contacts with more defects]{\label(SubFig2){subfig:jsm}
    \igr[height=63pt]{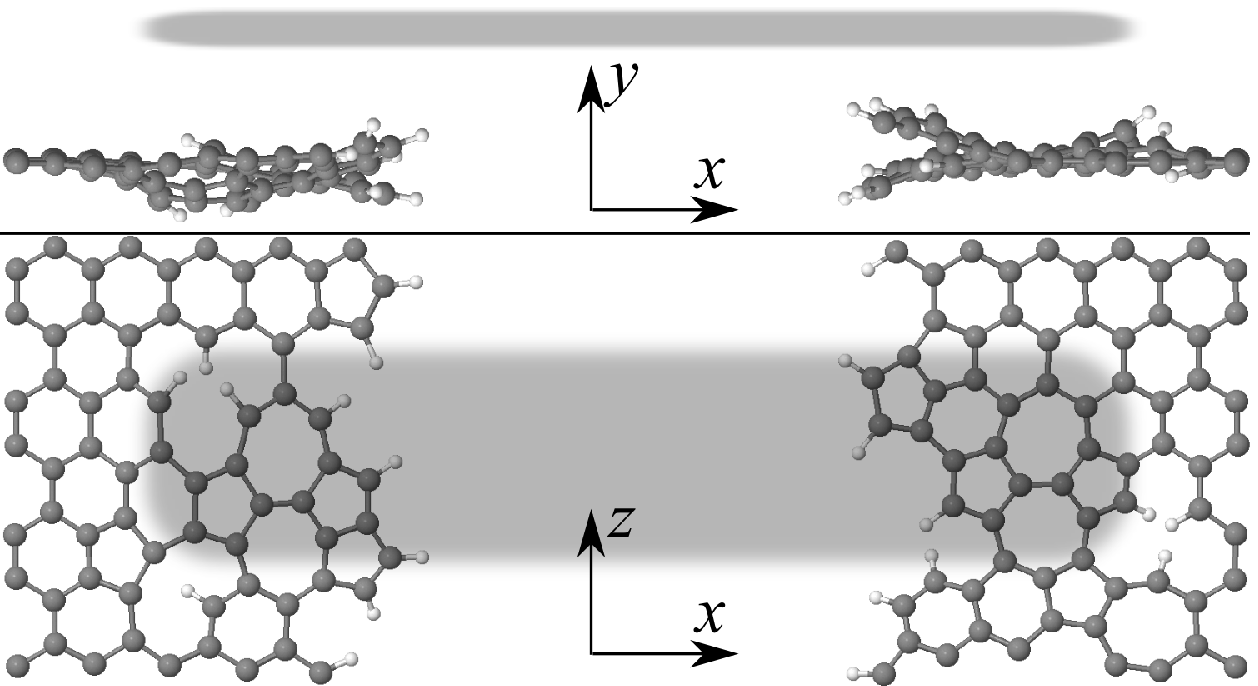}}
    \subfigure[\small\ Large contacts with more defects]{\label(SubFig2){subfig:jlm}
    \igr[height=63pt]{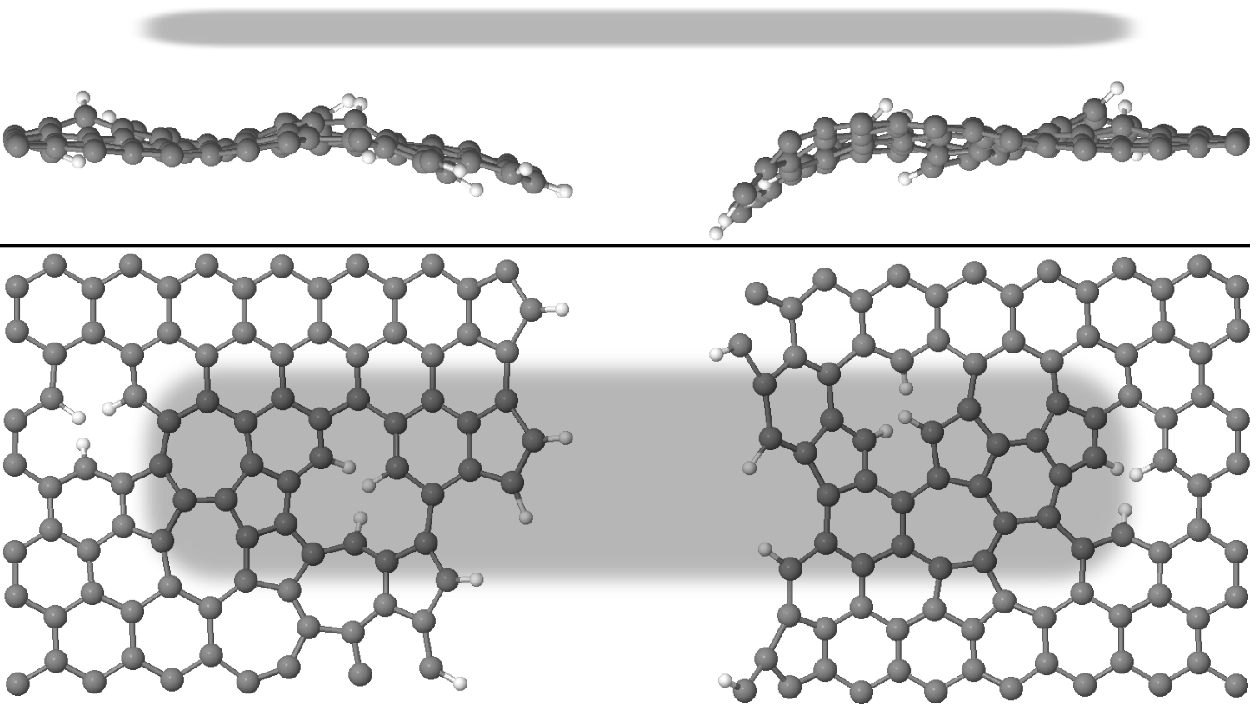}}
\caption{\small Projections of the contact area to the $x-z$ plane, viewed from above and from the side. The grey could represents the molecule above it. The gap and the middle section of the molecule are shrunk to fit the figure.}\label{fig:junctions}
\end{figure}
\begin{figure*}[!ht]
\centering
    \subfigure[\small\ $T(E)$ for the small contacts]{\label(SubFig1){subfig:ets}
    \igr[height=130pt]{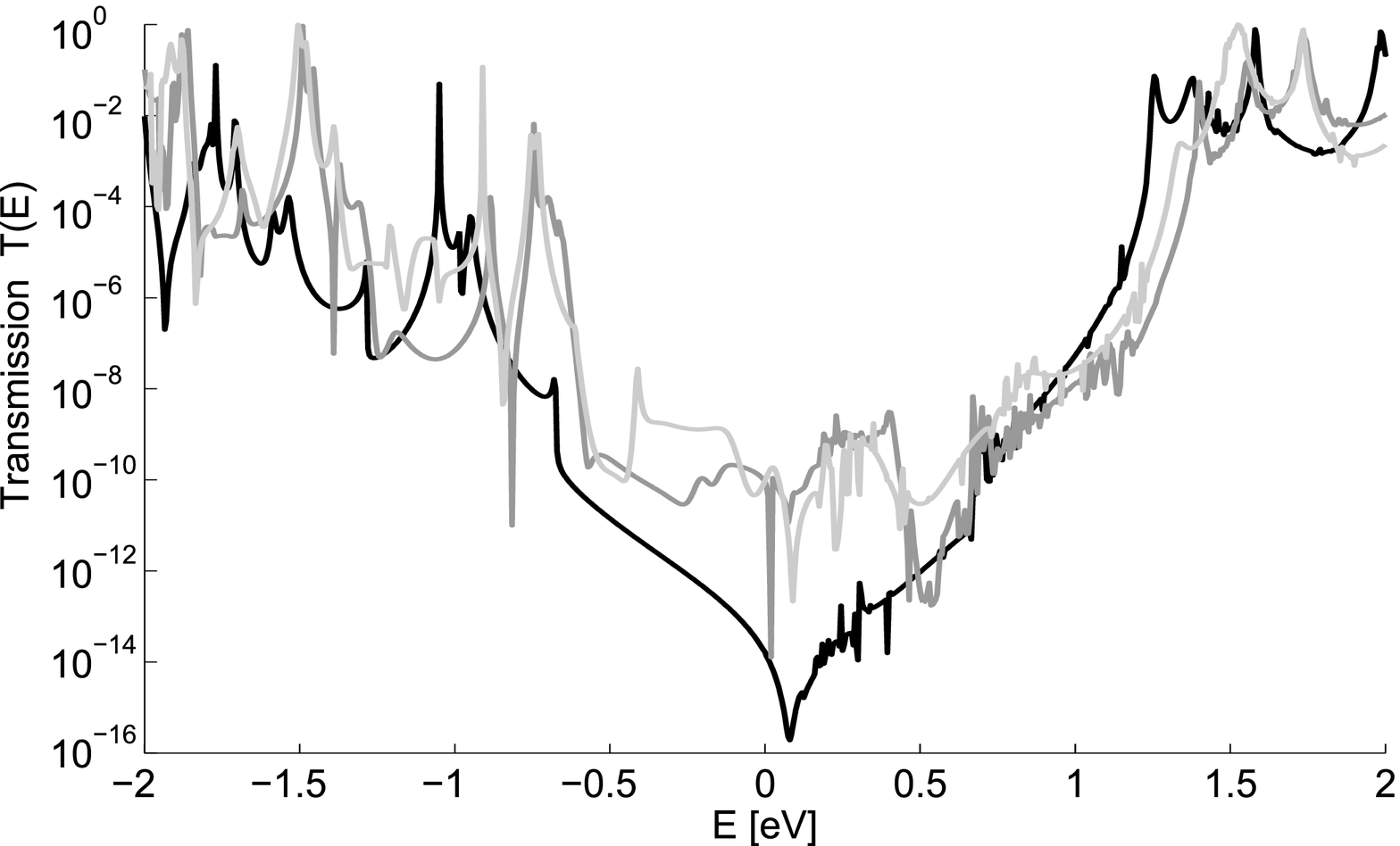}}
    \subfigure[\small\ $I(V)$ for the small contacts]{\label(SubFig1){subfig:bcs}
    \igr[height=130pt]{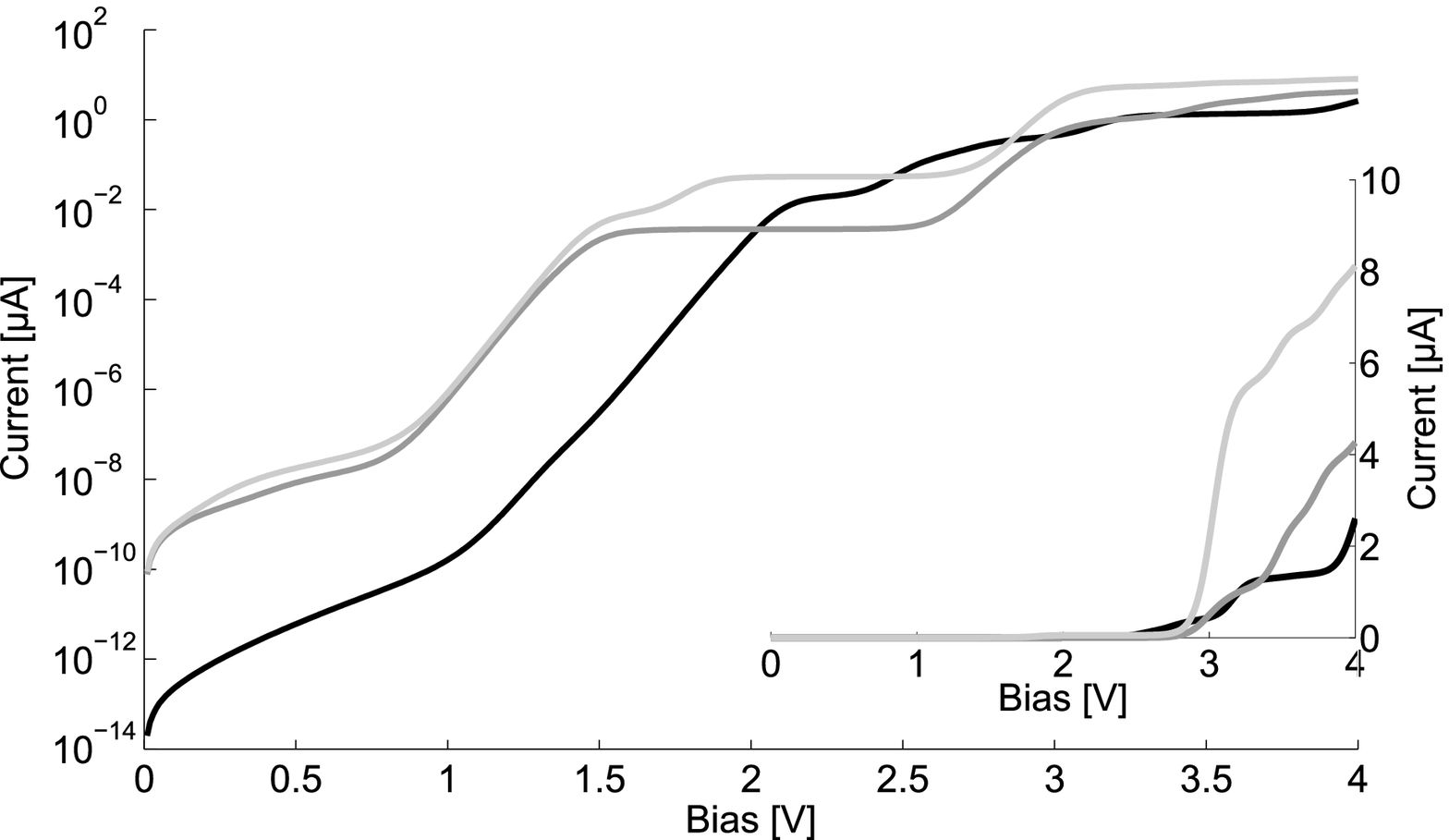}}
\\
    \subfigure[\small\ $T(E)$ for the large contacts]{\label(SubFig2){subfig:etl}
    \igr[height=130pt]{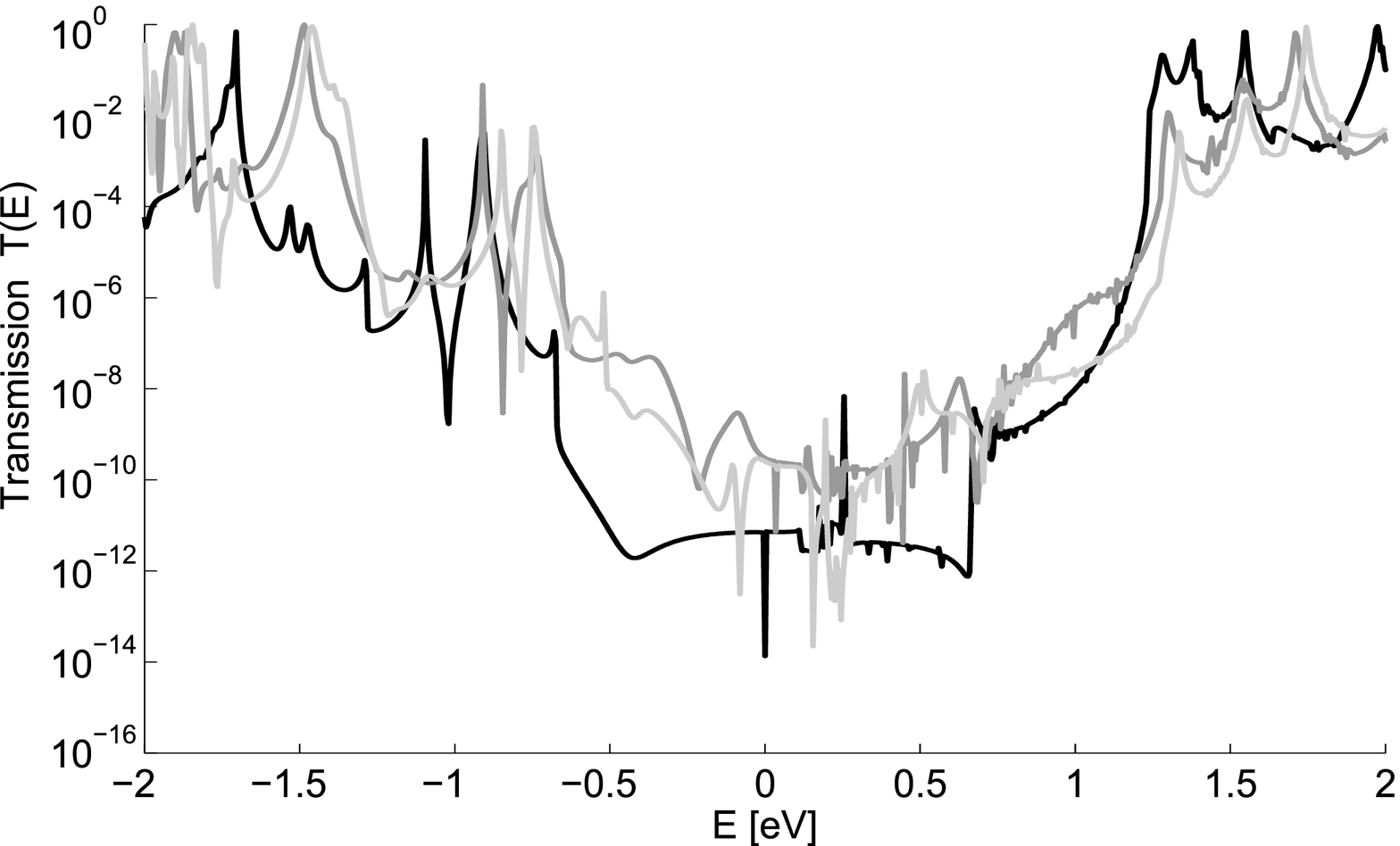}}
    \subfigure[\small\ $I(V)$ for the large contacts]{\label(SubFig2){subfig:bcl}
    \igr[height=130pt]{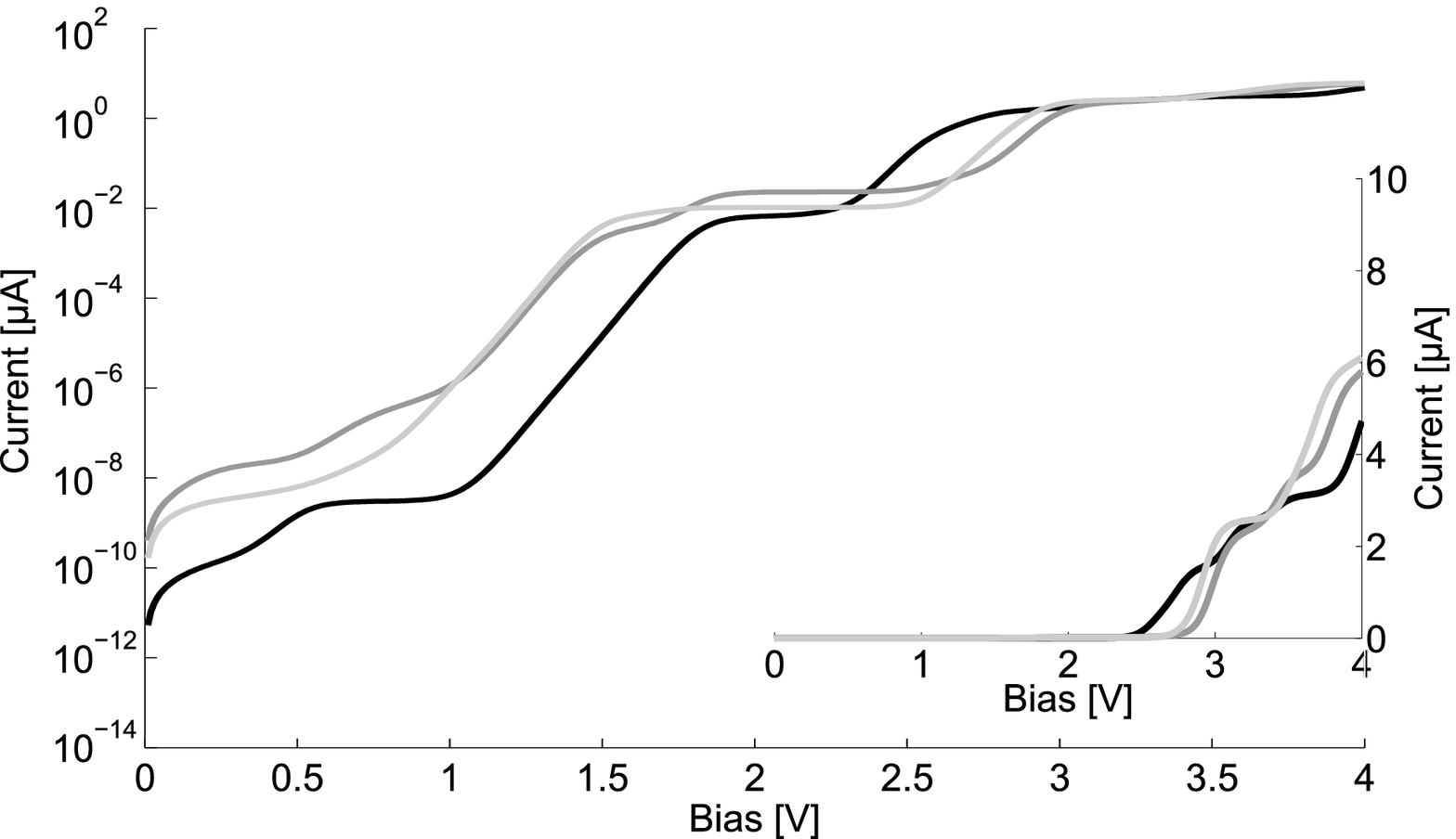}}
\caption{\small Transport probability as function of the energy
and room temperature current as function of the applied bias
voltage. The current is presented on a logarithmic scale and also
on a linear scale in the inset. Black curve corresponds to perfect
electrodes, dark gray to CNTs with a few defects and light gray to
CNTs with many defects. \mbox{Note the {\it threshold} voltage
below 3V.}}\label{fig:transport}
\end{figure*}
In the second stage, we double the concentration
of each kind of defect. In the presence of such defects, the
relaxed geometries of the CNT surfaces in the vicinity of the
anchor groupss are shown in Fig.~\ref{fig:junctions}, where for
clarity, both side and top views are shown and the attached
molecule is represented simply by grey shaded regions.

To demonstrate that fluctuations are reduced in the presence of
more extended anchor groups, we consider the case of 'small' and
'large' anchor groups. The small-anchor configuration is shown in
Fig.~\ref{fig:molecule}, where only the phenanthrene group is in
direct contact with the CNT. For the large anchor group, we
consider the case where the OPE spacers also overlay the CNT. For
the small and large anchors, attached to CNTs containing no
defects and two non-zero concentrations of defects, transmission
and $I(V)$ curves are shown in Fig.~\ref{fig:transport}.
Interestingly, above 3V, we find that lattice defects increase the
electrical conductance, by broadening the Breit-Wigner resonances
around $\pm1.5$V. This is particularly striking for the small
anchor groups.

To quantify the conductance fluctuations of the results presented in Fig.~\ref{subfig:bcs} and \ref{subfig:bcl}, at each voltage $V$, we compute the finite-voltage conductance $g(V)= I(V)/V$, for each of the three curves in Fig.~\ref{subfig:bcs} and \ref{subfig:bcl}. At each voltage, we then compute the mean and standard deviation $\sigma (V)$ of each set of three curves. For the small and large anchor groups, we denote the standard deviation $\sigma_{\textrm{small}} (V)$ and $\sigma_{\textrm{large}} (V)$ respectively. As a measure of their relative conductance fluctuations over a given voltage window, we define the ratio
\begin{equation}
\alpha = \left(\int_{V_1}^{V_2} \sigma_{\textrm{small}}(V)\textrm{d} V\right)\left/\left(\int_{V_1}^{V_2} \sigma_{\textrm{large}}(V)\textrm{d} V\right)\right..
\end{equation}
When integrating from 0 V up to 4 V, we find $\alpha = 3.7$, but when integrating only above the switch-on voltage from 3V to 4V, we find $\alpha =  5.6$. Both cases clearly demonstrate that the $I(V)$ curves with large anchor groups are far more resilient to lattice defects than those with smaller anchor groups.

In common with the measured 3-3 OPE
molecule~\cite{marquardt_electroluminescence_2010}, we find that
$I(V)$ curves of our 1-1 OPE are strongly non-linear, with a
threshold voltage for charge transport of $V_{CT}$, which supports
that this threshold voltage is an intrinsic feature of this class
of molecules. The 3-3 OPE threshold voltage was measured to be ca.
$V_{CT} = 2V$, whereas in our case the 1-1 OPE threshold is
predicted to be $V_{CT} = 2.5 - 3V$. In our calculations,
transport takes place by phase-coherent tunnelling and since the
Fermi energy is close to the centre of the HOMO-LUMO gap, $V_{CT}$
is approximately twice the value of the HOMO-LUMO gap, which is
larger for the smaller molecule.

\section{Conclusions}

Single-molecule electronics is an embryonic technology, whose development is hampered by conductance fluctuations due to atomic-scale variation at the contacts to electrodes. Using \textit{ab-initio} methods, we have demonstrated that conductance fluctuations of a single molecule bridging carbon-based electrodes can be significantly decreased by increasing the size of planar aromatic anchor groups, which $\pi-\pi$ bond to the electrodes. As a specific example, we have examined the  2,6-dibenzylamino core-substituted naphthalenediimide (NDI) chromophore attached to carbon nanotube electrodes, which is closely related to a molecule measured recently by Marquardt \textit{et al}.~\cite{marquardt_electroluminescence_2010} In agreement with their measurements, we find a non-linear $I(V)$ with strong current suppression at low voltages. The agreement between their threshold and our numerical value is of course a coincidence, because the molecule measured by Marquardt \textit{et al}~\cite{marquardt_electroluminescence_2010} is similar, but not identical to ours. Furthermore it is well known that DFT may not predict accurately the position of the Fermi energy relative to the HOMO and LUMO levels and we have neglected the possibility that the molecule can become charged. Nevertheless our calculations demonstrate that when the Fermi energy is located far from the HOMO or LUMO resonance peaks, a threshold voltage with a value close to the one measured by Marquardt \textit{et al}~\cite{marquardt_electroluminescence_2010} arises naturally within a phase-coherent tunnelling picture of transport, of the kind discussed in Ref.~\onlinecite{sedghi_long-range_2011}.

The transition from noble-metal-based electrodes to carbon-based electrodes represents a paradigm shift for molecular electronics. For the future, it will be fruitful to examine a range of alternative planar anchor groups, including pyrene and pyrene derivatives, which are known to bind strongly to graphene and CNTs,~\cite{g25}
larger polycyclic aromatics and molecules terminated by multiple anchor groups.

\section*{Acknowledgement}
We gratefully acknowledge discussions with Iain Grace and funding from the Marie-Curie ITNs FUNMOLS and NANOCTM. Funding is also provided by EPSRC and METRC.

\section*{Supporting information}
The relaxed coordinates of the CNT-molecule-CNT systems with lattice defects, together with the corresponding 3D rotatable images are available at \texttt{peterfalvi.web.elte.hu/Bulky.html}.




\end{document}